\documentclass[a4paper,11pt]{article}
\usepackage{pos}
\usepackage{subcaption}
\usepackage{epsfig}
\usepackage{color} 
\usepackage{xcolor}
\usepackage{bm}
\usepackage{afterpage}

\title{Charmonium-like resonances in coupled   $D\bar D$ - $D_s \bar D_s$ scattering}

\author*[a,b,c]{S. Prelovsek}
\author[a]{S. Collins}
\author[d]{D. Mohler}
\author[d,e]{M. Padmanath}
\author[a]{S. Piemonte}

\affiliation[a]{Institute for Theoretical Physics, University of Regensburg, 93040 Regensburg, Germany  }
\affiliation[b]{Faculty of Mathematics and Physics, University of Ljubljana, Slovenia}
\affiliation[c]{Jozef Stefan Institute, Ljubljana, Slovenia}
\affiliation[d]{GSI Helmholtzzentrum f\"ur Schwerionenforschung, 64291 Darmstadt, Germany}
\affiliation[e]{Helmholtz-Institut Mainz, Johannes Gutenberg-Universit\"at, D-55099 Mainz, Germany}

 \emailAdd{sasa.prelovsek@ijs.si}
\emailAdd{pmadanag@uni-mainz.de} 

\abstract{Charmonium-like resonances and bound states with isospin zero and $J^{PC}=0^{++},~1^{--},~2^{++},~3^{--}$ are extracted on the lattice.  Coupled  $D\bar D$ and $D_s\bar D_s$ scattering suggests  three charmonium-like states with $J^{PC}=0^{++}$  in addition to $\chi_{c0}(1P)$: a so far unobserved $D\bar D$ bound state just below threshold, a conventional  resonance likely related to $\chi_{c0}(3860)/\chi_{c0}(2P)$ and a narrow  resonance just below the $D_s\bar D_s$ threshold with a large coupling to $D_s\bar D_s$ likely related to $X(3915)/\chi_{c0}(3930)$. One-channel $D\bar D$ scattering  renders   resonances and bound states with $J^{PC}= 1^{--},~2^{++},~3^{--}$ related to the observed conventional charmonia. Lattice QCD ensembles from the CLS consortium with $m_{\pi}\simeq 280$ MeV are utilized. }

\FullConference{%
 The 38th International Symposium on Lattice Field Theory, LATTICE2021
  26th-30th July, 2021
  Zoom/Gather@Massachusetts Institute of Technology
}

\begin{document}
\maketitle

\section{Introduction}

\begin{figure}[thb]
	\begin{center}
	  \centerline{\includegraphics[width=0.7\textheight]{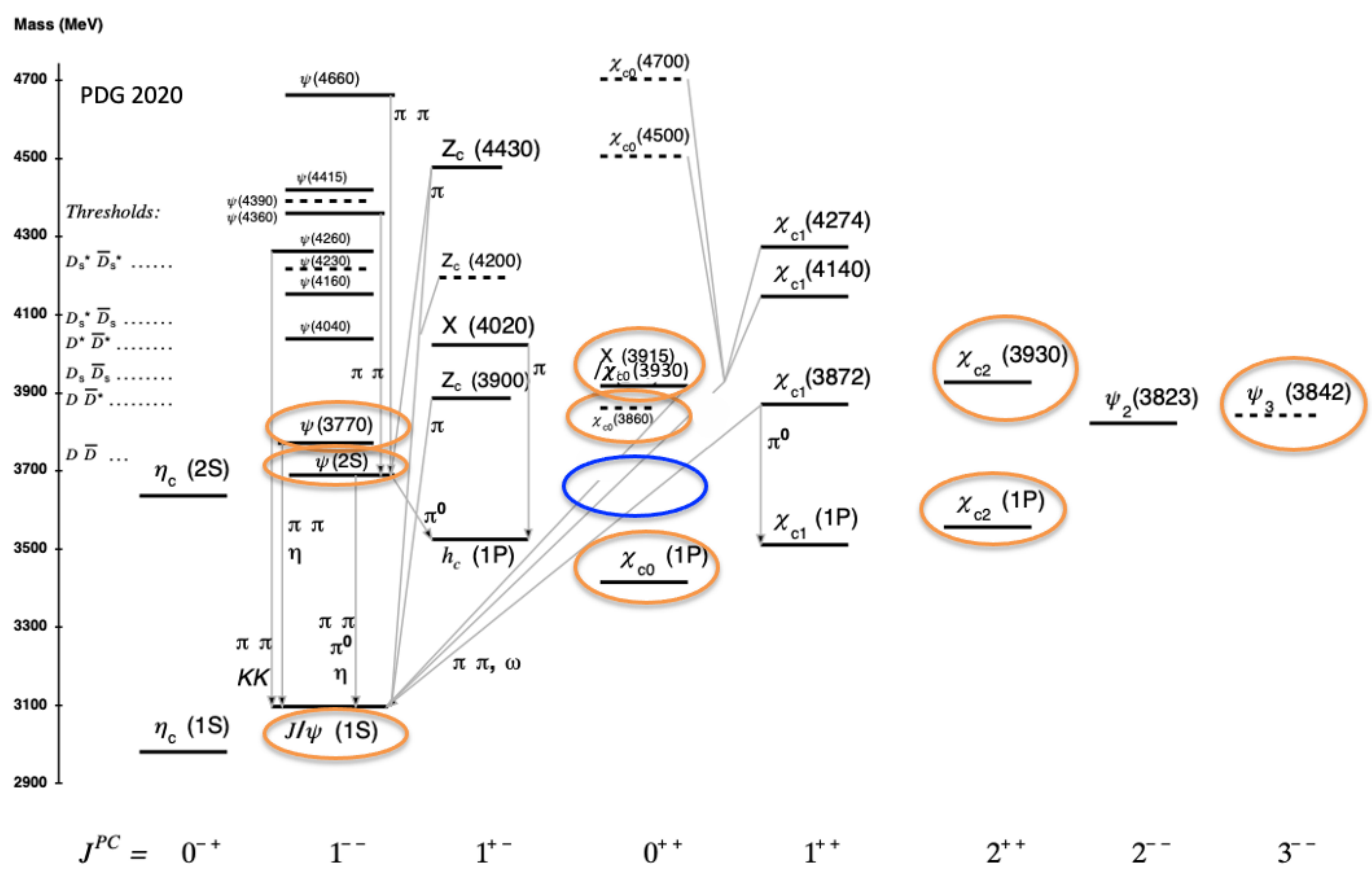}     }        
		\caption{  Charmonium-like  states from PDG \cite{pdg} are shown in black. The ellipses indicate   states considered in the present lattice study: $J^{PC}=0^{++},~2^{++}$ and $J^{PC}=1^{--},~3^{--}$  were extracted in \cite{Prelovsek:2020eiw} and \cite{Piemonte:2019cbi} respectively. The  state indicated in blue is the lattice prediction of yet-unobserved unconventional state that couples strongly to $D\bar D$ and  lies near this threshold. Other states have been experimentally discovered: all of them seem to be conventional charmonia, except for $X(3915)/\chi_{c0}(3930)$ which is  unconventional state that couples strongly to $D_s\bar D_s$  according to this lattice study.  }
	\label{fig:states-considered}
	\end{center}
  \end{figure}

 We study charmonium-like resonances with isospin zero by simulating the scattering on the lattice.  
 Studies of charmonium-like states are of particular interest as most of the exotic hadrons discovered contain a $\bar cc$ pair. Furthermore, all these exotic hadrons are resonances and can decay strongly.     
  We consider the resonances and bound states with $J^{PC}=0^{++},~1^{--},~2^{++},~3^{--}$ indicated in Fig. \ref{fig:states-considered}, where the relevant thresholds are also shown. The scalars are extracted from the coupled channel scattering $D\bar D$ - $D_s \bar D_s$ since two resonances reside near the $D_s \bar D_s$ threshold. The negative parity states of interest lie significantly below the $D_s \bar D_s$ threshold and are explored  with one-channel $D\bar D$ scattering. The spin-two resonance is also extracted in the approximation of one-channel $D\bar D$ scattering. Details   are provided in publications \cite{Piemonte:2019cbi} and \cite{Prelovsek:2020eiw} which consider the negative and positive parity sectors, respectively. 
 
  \section{Extracting scattering amplitudes from lattice}
  
  The infinite-volume scattering matrices $t_{ij}(E_{cm})$ 
  are determined from the finite-volume eigen-energies via L\"uscher's formalism.  The eigen-energies are obtained from the correlation matrices based on a number of $\bar cc$, $D\bar D$, $D_s \bar D_s$ and $J/\psi \omega$ interpolating fields with appropriate quantum numbers. The channel  $J/\psi \omega$ is treated as decoupled, while the 
  channel $\eta_c \eta$ is omitted from the simulation.  
    We consider  the systems with total momenta $|\vec P|=0,~2\pi/L,~ \sqrt{2} (2\pi/L)$ on lattices with $L=24a,~32a$ in order to constrain the scattering matrix at more values of the center-of-momentum energy $E_{cm}$. The $c\bar c$ annihilation is  omitted like in most of the  charmonium studies on the lattice, while  other Wick contractions as evaluated using the distillation method. 
    The simulation is performed on the $N_{f}=2+1$ ensembles U101 and H105 generated by the CLS consortium \cite{Bruno:2014jqa}  with $a=0.08636(98)(40)$~fm and unphysical $m_{u/d}>m_{u/d}^{exp}$ and $m_{s}<m_{s}^{exp}$, such that $2m_{u/d}+m_s$ is close to the physical value.  The value of the charm quark mass (related to $\kappa_c=0.12315$) is  slightly higher than physical and the resulting masses of the relevant stable hadrons are  collected in Table \ref{tab:ensemble}.  In particular, the $D\bar D$ and $D_s\bar D_s$ thresholds are closer  to each other than in experiment as shown in Fig. \ref{fig:masses}. 
   
   The energy dependence of scattering matrices $t_{ij}(E_{cm})$  is parametrized and the parameters are fitted as to best satisfy the  L\"uscher quantization condition for all lattice eigen-energies simultaneously.  The {\it TwoHadronsInBox} package  \cite{Morningstar:2017spu} is utilised for this.  The typical parametrizations used for the real part of the inverse scattering amplitude are of Breit-Wigner type $\tilde K^{-1}/E_{cm}=p^{2l+1} \cot\delta /E_{cm}=a+b E_{cm}^2$ and $\tilde K_{ij}^{-1}/E_{cm}= a_{ij}+b_{ij} E_{cm}^2$ for one-channel and two-channel scattering, respectively, with more details given in  \cite{Prelovsek:2020eiw,Piemonte:2019cbi}.     
   The study makes several simplifying assumptions (detailed
in Section 5 of \cite{Prelovsek:2020eiw}) necessary for a first investigation of this coupled-channel system.

  \begin{table}[tb]
  \begin{center}
    \begin{tabular}{cccccc}
      \hline
  $m_\pi$~[MeV]& $m_K$~[MeV] & $m_D$~[MeV] & $m_{D^*}$~[MeV] & $m_{D_s}$~[MeV] & $M_{av}$~[MeV]\\
      \hline
      \hline
      280(3) & 467(2) & 1927(2) & 2050(2) & 1981(1) & 3103(3)\\
      \hline
  \end{tabular}
  \end{center}
  \caption{Hadron masses   for the gauge
    configurations used in this study, where $M_{av}=(m_{\eta_c}+3m_{J/\psi})/4$. }
  \label{tab:ensemble}
\end{table}

    \begin{figure}[tb]
	\begin{center}
	\includegraphics[width=0.55\textwidth]{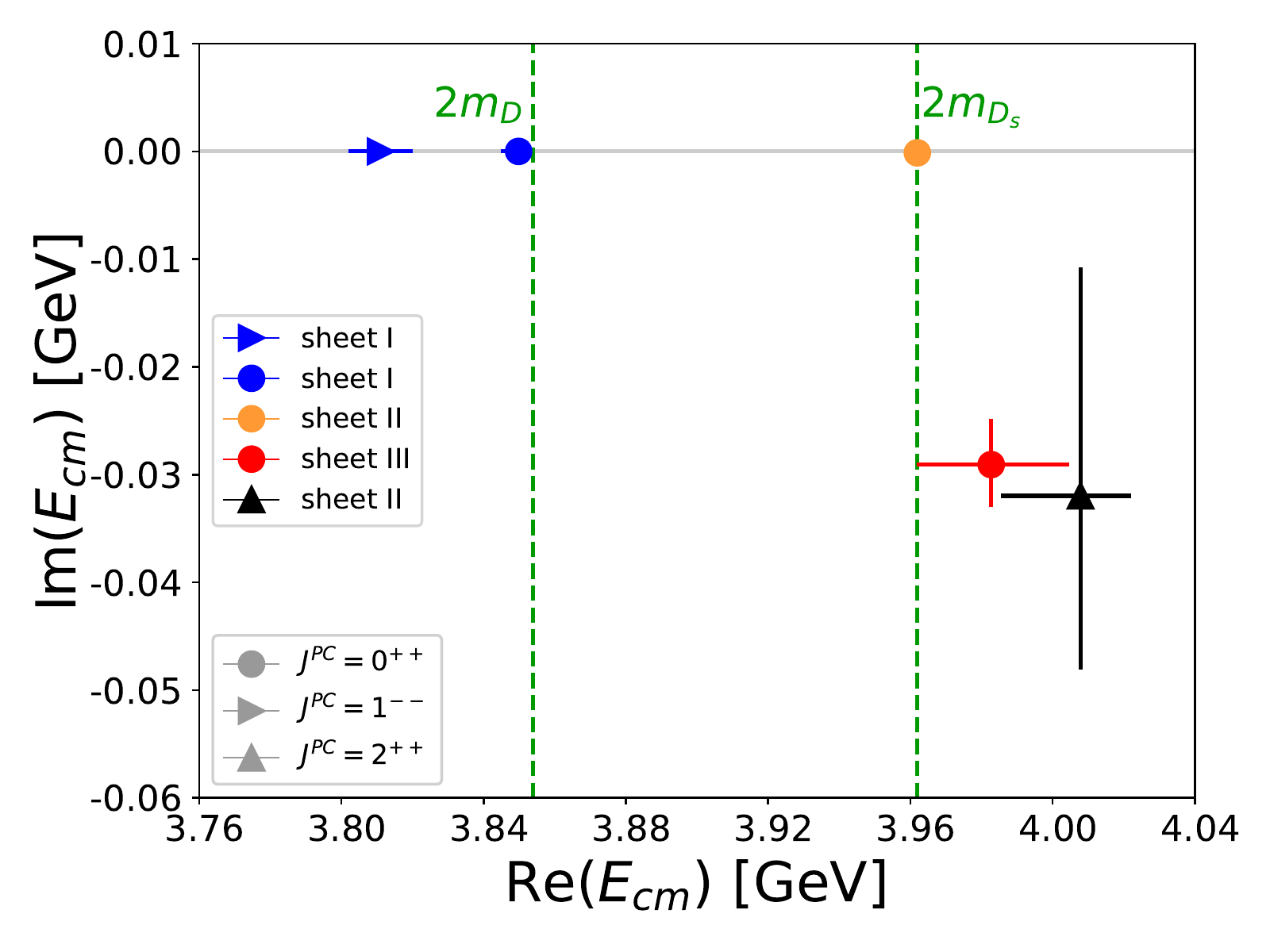}   
		\caption{ 
		  Pole singularities of the scattering amplitude/matrix in the complex energy plane that are extracted from this lattice study. These singularities   are  associated with the  charmonium-like states and various symbol shapes denote the corresponding  $J^{PC}$. The $3^{--}$ state is not presented as we  did not extract its width.  }
	\label{fig:summary-poles}
	\end{center}
   \end{figure}

   \begin{figure}[p]
\centering
\begin{subfigure}{0.75\textwidth}
\includegraphics[width=\textwidth]{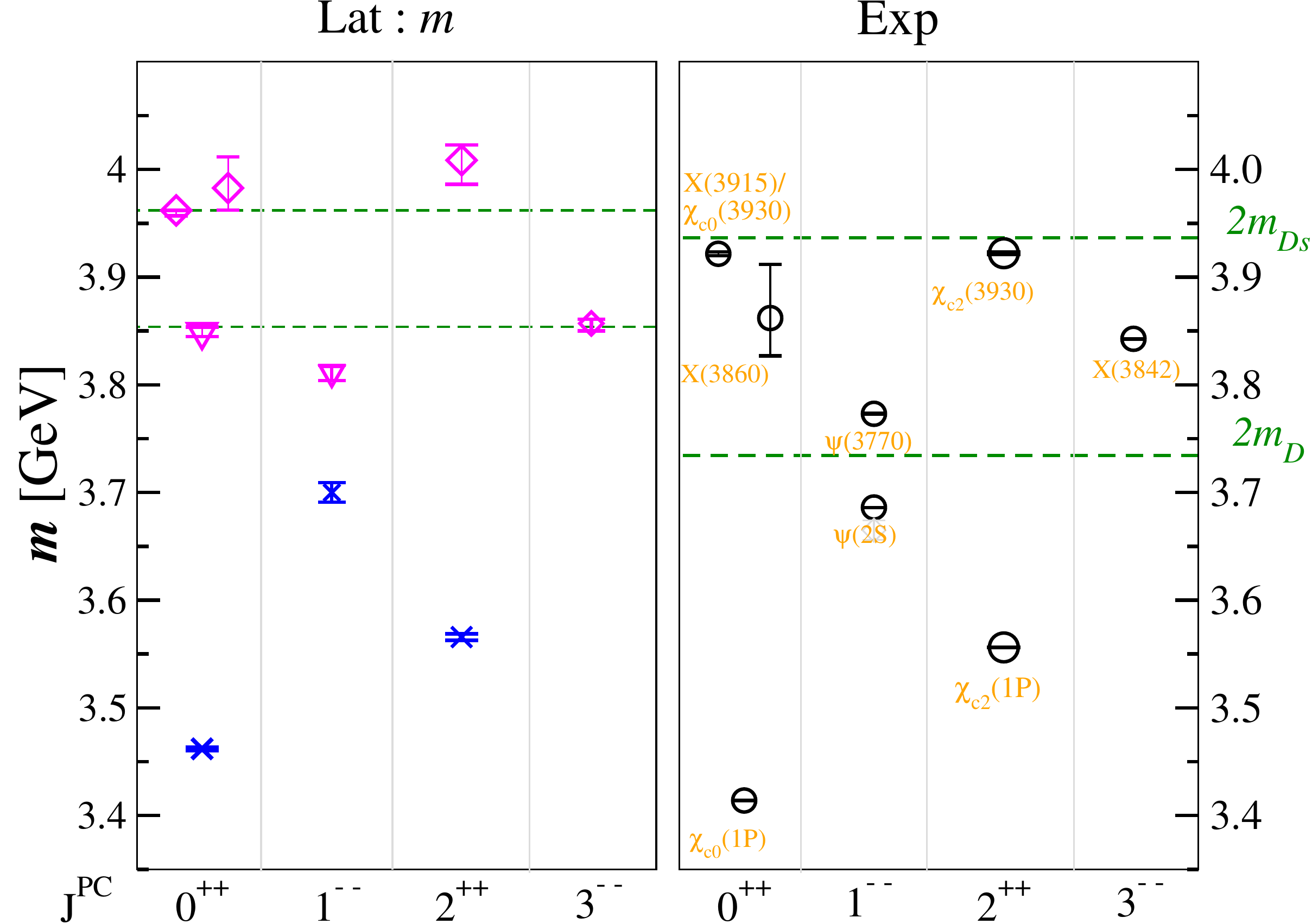}
\caption{ Left pane: The masses $m$   obtained from our lattice study with unphysical quark masses $m_{u/d}>m_{u/d}^{exp}$, $m_{s}<m_{s}^{exp}$ and $m_c\gtrsim m_c^{exp}$. The green lines denote positions of the $D\bar D$ and $D_s\bar D_s$ thresholds on our lattice with   $m_D \simeq 1927~$MeV and $m_{Ds}\simeq 1981~$MeV. } \label{fig:a}
\end{subfigure} 

\vspace{0.3cm}

\begin{subfigure}{0.75\textwidth}
\includegraphics[width=\textwidth]{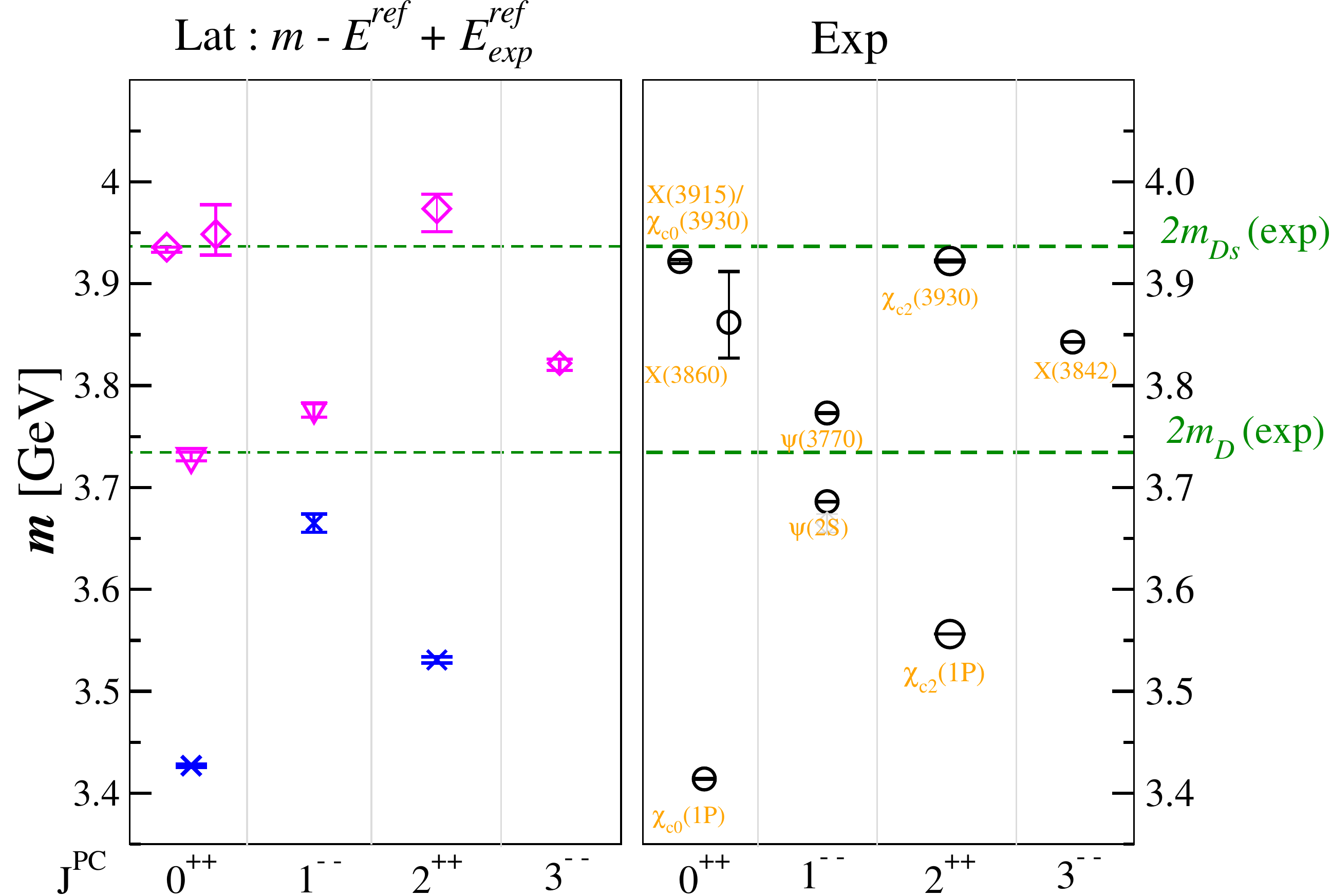}
\caption{  Left pane:   The same masses $m$ as above, but shifted to $m-E^{ref}+E^{ref}_{exp}$ in order to account for the dominant effect of unphysical quark masses in the simulation.  The reference energy   is $E^{ref}=2m_D$ ($2m_{D_s}$) for the state closest
                    to the $D\bar D$ ($D_s\bar D_s$) threshold, while
                    $E^{ref}=M_{av}=\tfrac{1}{4}(3m_{J/\psi}+m_{\eta_c})$  for
                    the remaining four states. The green lines denote experimental thresholds.} \label{fig:b}
\end{subfigure}
\caption{    Masses  of  charmonium-like states with isospin zero  from a lattice simulation \cite{Prelovsek:2020eiw,Piemonte:2019cbi} (left) compared to
                   experiment (right).    Left: The magenta symbols  correspond
                    to hadrons extracted via the scattering analysis on the lattice:
                    diamonds  represent resonances and triangles represent
                    bound states. The blue crosses are extracted directly from
                    the lattice energies.   Right:  experimental spectrum from the PDG \cite{pdg}, where $\chi_{c0}(3930)$   \cite{chic03930} and $X(3915)$  are now identified as the same state. }
\label{fig:masses}
\end{figure}

  \section{Summary of the resulting hadrons and their relation to  experiment}    \label{sec:summary}

    Let us summarize the properties of the charmonium-like hadrons
    found in this simulation. All results were obtained at unphysical quark masses and are not extrapolated  to the continuum limit.    
     The resonances and bound states are related to the poles of the scattering matrix.  
    The  locations of the poles   in the complex energy plane     are given in Fig.~\ref{fig:summary-poles}.  The masses of 
 the  charmonium-like states correspond to $\mathrm{Re}(E_{cm}^{pole})$ and are compared to the experimental spectrum in Fig.~\ref{fig:masses}. The resonance decay widths $\Gamma$ are given by $2\mathrm{ Im}(E_{cm}^{pole})$. We don't compare them directly to experiment since they depend on the phase space and therefore on  the position of the
 threshold, which  is different in the simulation.  Figure \ref{fig:g}  shows the coupling $g$ that parametrizes the full width of a  resonance  and compares it to experiment
    \begin{equation}
  \label{references}  
  \Gamma \equiv g^2   p_D^{2l+1}/m^2~\quad \mathrm{with}\quad l=0,1,2,3\quad  \mathrm{for} \ J^{PC}=0^{++},1^{--},2^{++},3^{--}~.
\end{equation}

Due to the unphysical quark masses   in the simulation, the   difference $m-E^{ref}$ is compared to experiment below and $m-E^{ref}+E^{ref}_{exp}$ is shown in Fig.~\ref{fig:masses}b. The reference
 energy $E^{ref}$ is   the  spin-averaged charmonium mass
 $M_{av}=\tfrac{1}{4}(3m_{J/\psi}+m_{\eta_c})$ for conventional candidates and a nearby threshold for states dominated by $D_{(s)}\bar D_{(s)}$.  
 
 \vspace{0.2cm}
 
 Below we review the spectroscopic properties of the  charmonium-like states, which are summarized also in Figs. \ref{fig:summary-poles}, \ref{fig:masses} and \ref{fig:g}:
 
\begin{figure}[h!]
	\begin{center}
	\includegraphics[width=0.5\textwidth]{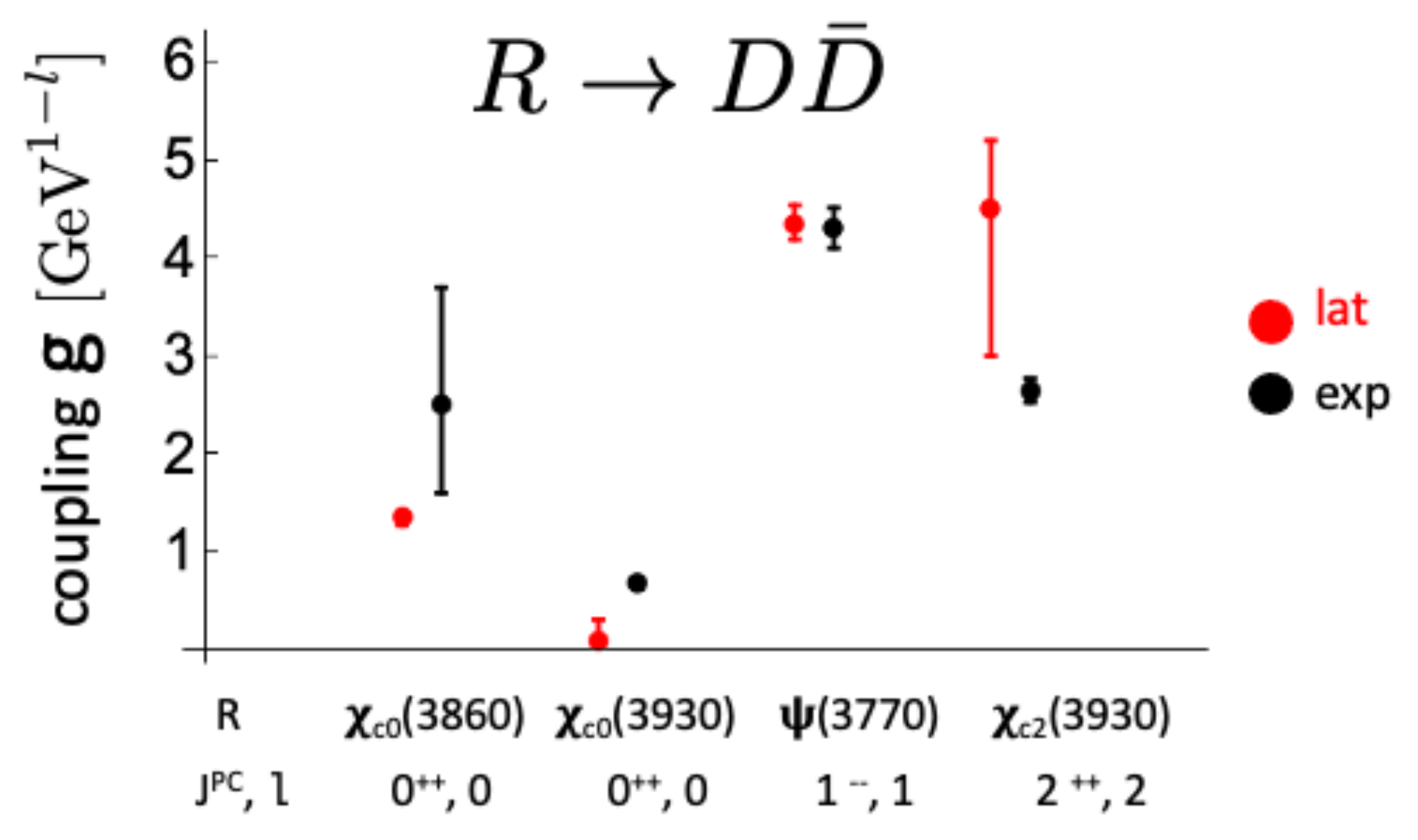}    
		\caption{    The coupling $g$ that parametrizes the decay width for various resonances $R\to D\bar D$ via  $\Gamma \equiv g^2   p_D^{2l+1}/m^2$ from this lattice simulation and from experiment.  The quantum numbers of $R$ and partial waves $l$ in the decay $R\to D\bar D$ are provided at the bottom.   }
	\label{fig:g}
	\end{center}
\end{figure}  

   \begin{itemize}
\item{ \bf{Ground states $\bm{ J/\psi,~ \psi(2S),~\chi_{c0}(1P) ,~ \bm{\chi_{c2}(1P)}}$}  }
    
      These states lie significantly below the $D\bar D$ threshold and their
      masses are extracted from the energies $m\!=\!E(P\!=\!0)$, resulting in a reasonable agreement with experimental masses.  
 
\item{\bf{$\bm{0^{++}}$ state dominated by $\mathbf{D \bar D}$ slightly below threshold } }

 The $D\bar D$ scattering near threshold in Fig. \ref{fig:DD} indicates the presence of a shallow $D\bar D$ bound state   with  the binding energy $m-2~m_D= -4.0 ^{~ + 3.7}_{~ -5.0}$~MeV.   Our   results  suggest  this state is likely  not a conventional  charmonium $\bar cc$, but owes its existence to a large
interaction in the $D\bar D$ channel near threshold. Such a state has not been  claimed by experiments (yet).  It would feature as a sharp increase of the $D\bar D$ rate just above the threshold if it has a  binding energy of a few MeV.   Various strategies for its experimental search   were proposed by E. Oset et al. and are listed in   Section 8.1 of \cite{Prelovsek:2020eiw}.

\begin{figure}[h!]
	\begin{center}
	\includegraphics[width=0.41\textwidth]{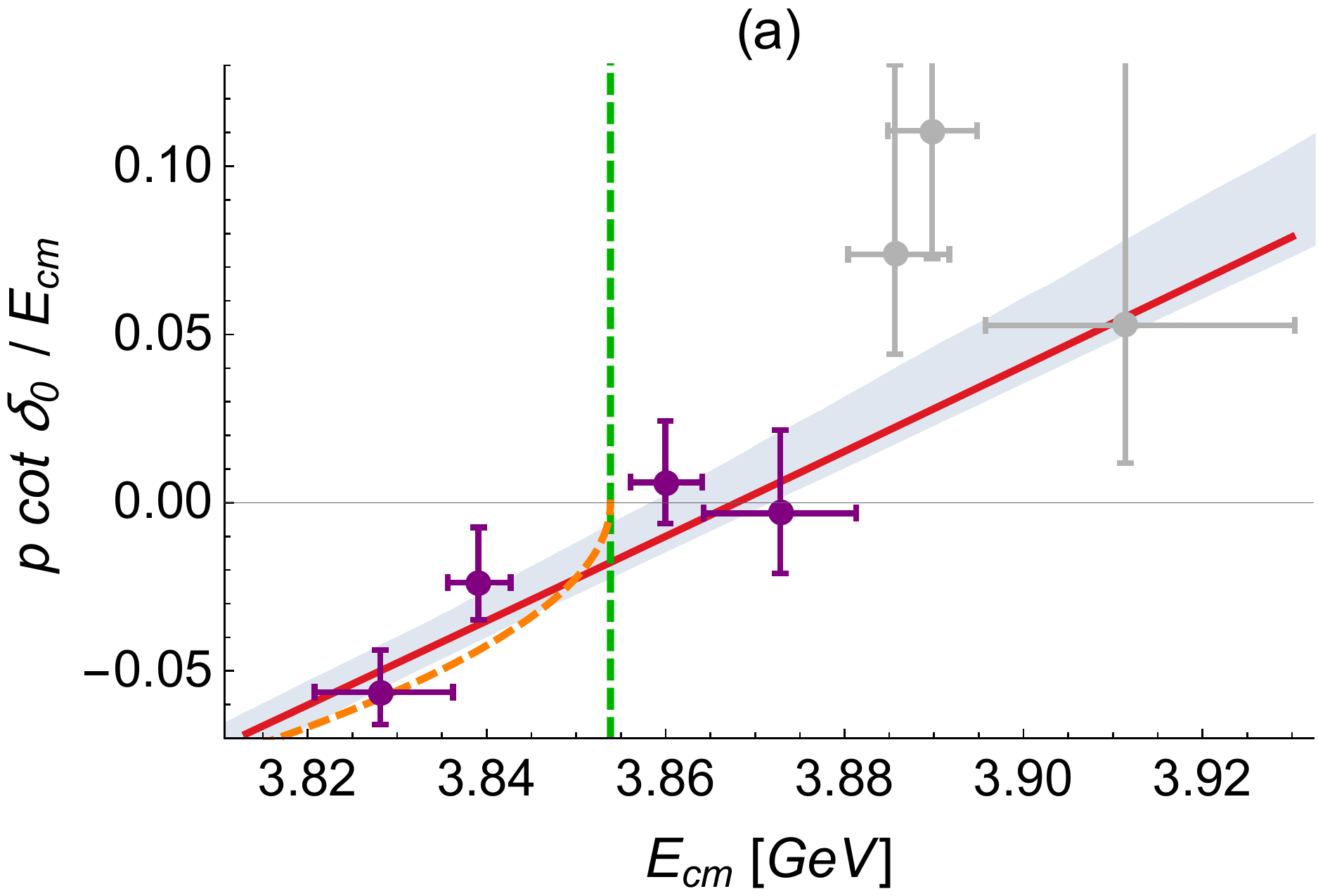} $\qquad$ 
		 \includegraphics[width=0.41\textwidth]{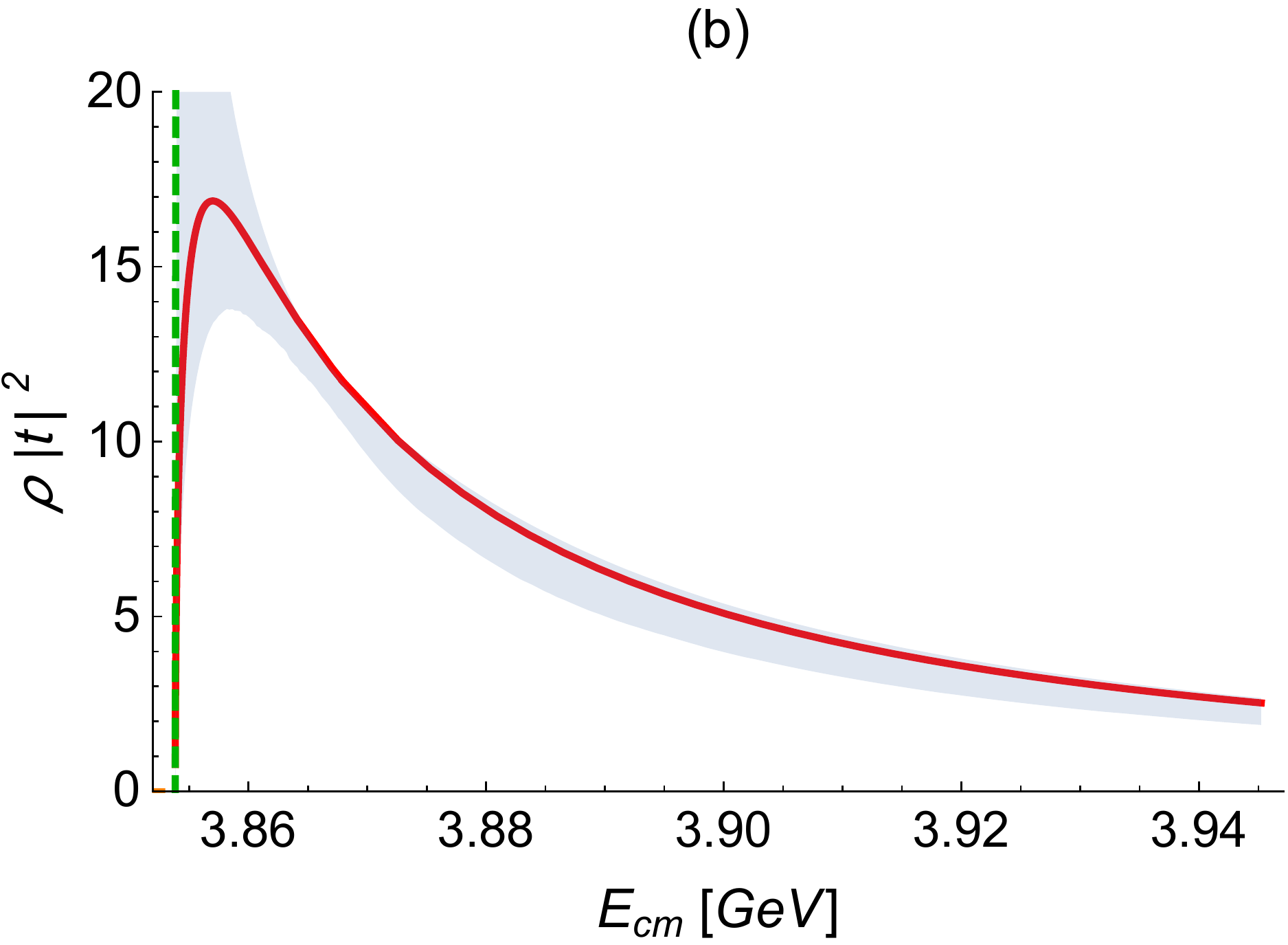}
		\caption{  $D\bar D$ scattering in partial wave $l\!=\!0$, where the threshold position is shown in green. Left: The bound state occurs at the energy where the red and orange curves intersect. Right: The shallow bound state is responsible for a sharp rise of the  $D\bar D$  rate $N_{D\bar D} \propto p\sigma\propto \rho |t|^2$ ($\rho=2p/E_{cm}$  and $t$ is the scattering  amplitude).   }
	\label{fig:DD}
	\end{center}
\end{figure}  

   The existence of a shallow $D\bar D$ bound state dubbed $X(3720)$
  was already suggested by an effective
   phenomenological model  based on the exchanges of light pseudoscalar and vector mesons   \cite{Gamermann:2006nm}\footnote{This state with
     $m\simeq 3.718~$GeV is listed in Table 4 of Ref. \cite{Gamermann:2006nm}.}. 
        In a molecular picture, a $0^{++}$ state is expected as a partner of $X(3872)$  via
   heavy-quark symmetry   \cite{Hidalgo-Duque:2013pva,Baru:2016iwj,Dong:2021juy}.
   A similar virtual 
     bound state with a binding energy of 20 MeV follows from the data of 
     the only previous lattice simulation of $D\bar D$ scattering 
     \cite{Lang:2015sba}\footnote{The presence of this state was not mentioned in Ref.~\cite{Lang:2015sba}, as 
     such virtual bound states were not searched for.}.

\item{\bf{$\bm{2^{++}}$ resonance and   relation to   $\bm{\chi_{c2}(3930)}$}}
   
   We find a resonance with $J^{PC}\!=\!2^{++}$  $D\bar
   D$ scattering with $l=2$  that is most likely related to the    conventional $\chi_{c2}(3930)$  aka $\chi_{c2}(2P)$   discovered by Belle \cite{Uehara:2005qd}
\begin{align} 
\label{dwave}
 \mathrm{lat}:\ \ &m-M_{av}=904 ^{~+14} _{~-22}~\mathrm{MeV}~,\quad g= 4.5 ^{+0.7}_{-1.5} \ \mathrm{GeV^{-1}}~\\
 \mathrm{exp} \ \chi_{c2}(3930):\ \ & m -M_{av}=854 \pm 1 ~\mathrm{MeV}~,\quad g =2.65 \pm 0.12 ~\mathrm{GeV^{-1}}~.\nonumber
\end{align} 
  The masses are
reasonably close, while  the coupling from lattice QCD is  larger than that measured
experiment, but  not inconsistent given the  statistical and systematic   uncertainties. 

\item{\bf{Broad $\bm{0^{++}}$ resonance  and  relation to  $\bm{\chi_{c0}(3860)}$}}
      
     The coupled   $D\bar D-D_s\bar D_s$ scattering shows a broad peak in the $D\bar D$ channel featured in the left pane in Fig. \ref{fig:DD-DsDs-with2-t}). It is related to a   resonance that couples mostly to $D\bar D$. We compare it   to    $\chi_{c0}(3860)$   discovered by
Belle \cite{Chilikin:2017evr}  that is a candidate for the conventional $\chi_{c0}(2P)$
     \begin{align}
   \label{chic0prime } 
  \mathrm{lat}: \ \ &m-M_{av}=880  ^{+28}_{-20} ~\mathrm{MeV}\ ,\quad g = 1.35~^{+0.04}_{-0.08}   \ \mathrm{GeV}~\\
  \mathrm{exp}\ \chi_{c0}(3860):\    &m-M_{av}=793~^{+48}_{-35} ~\mathrm{MeV}\ , \quad g=2.5 ^{+1.2}_{-0.9} ~\mathrm{GeV}~.\nonumber
 \end{align}
The mass and coupling are reasonably consistent with experiment, in particular, when considering the experimental errors and the
systematic uncertainties in the lattice results. 
    
     \begin{figure}[h!]
	\begin{center}
	\includegraphics[width=0.32\textwidth]{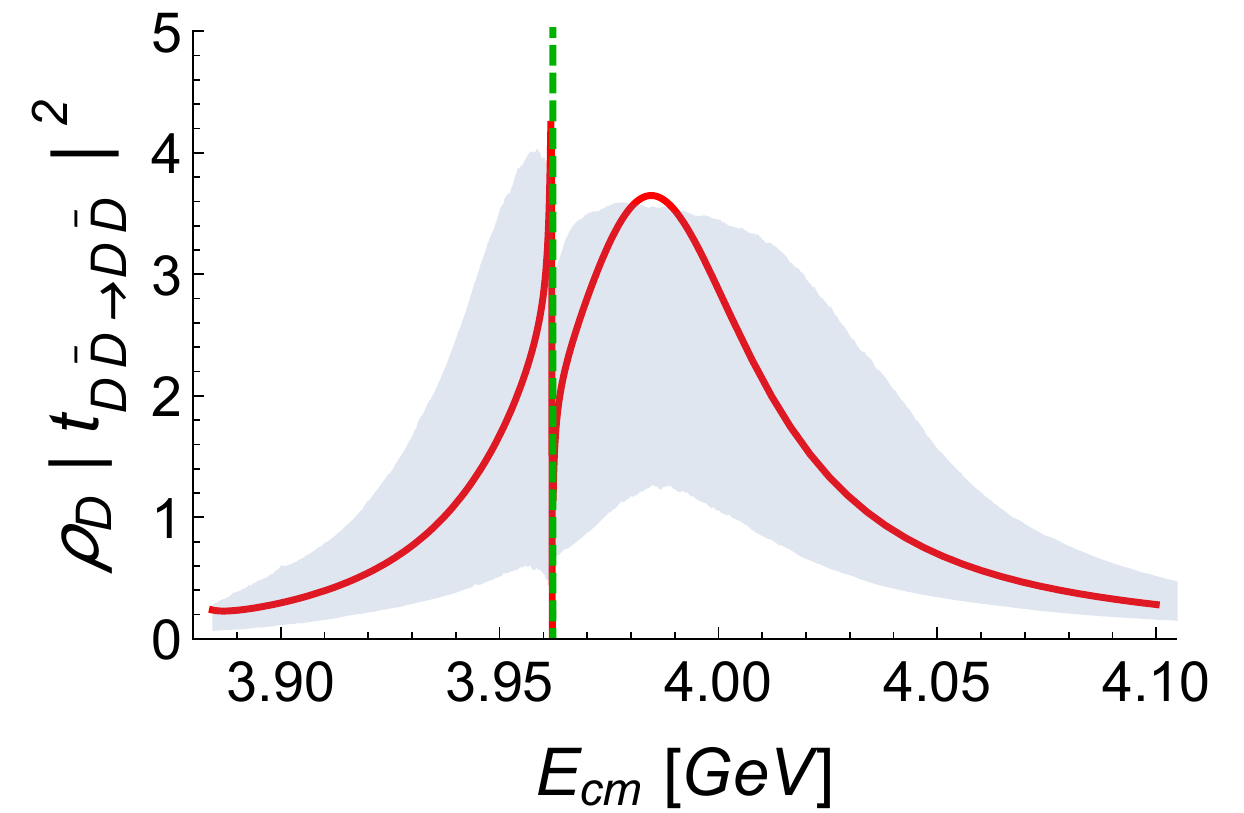}   
	\includegraphics[width=0.33\textwidth]{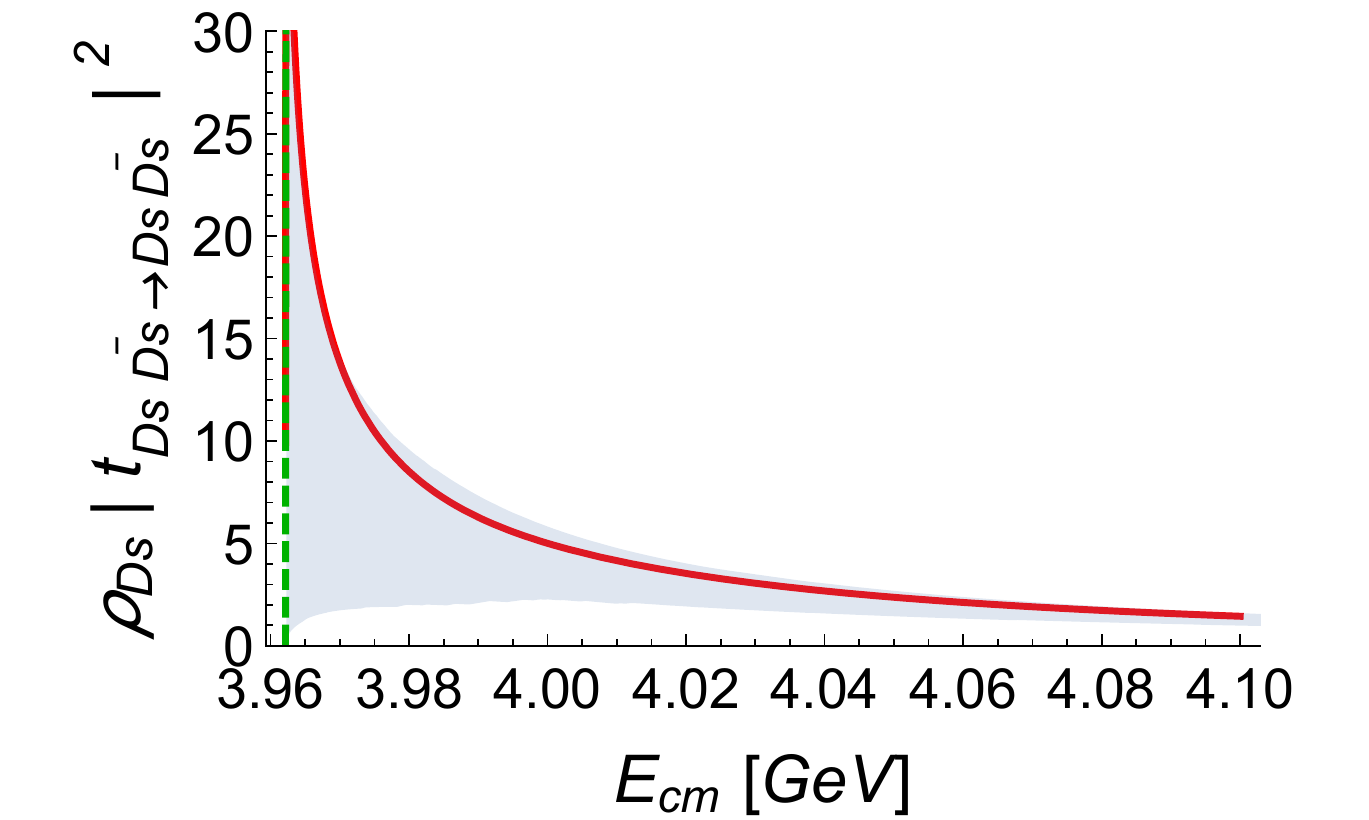}   
	\includegraphics[width=0.33\textwidth]{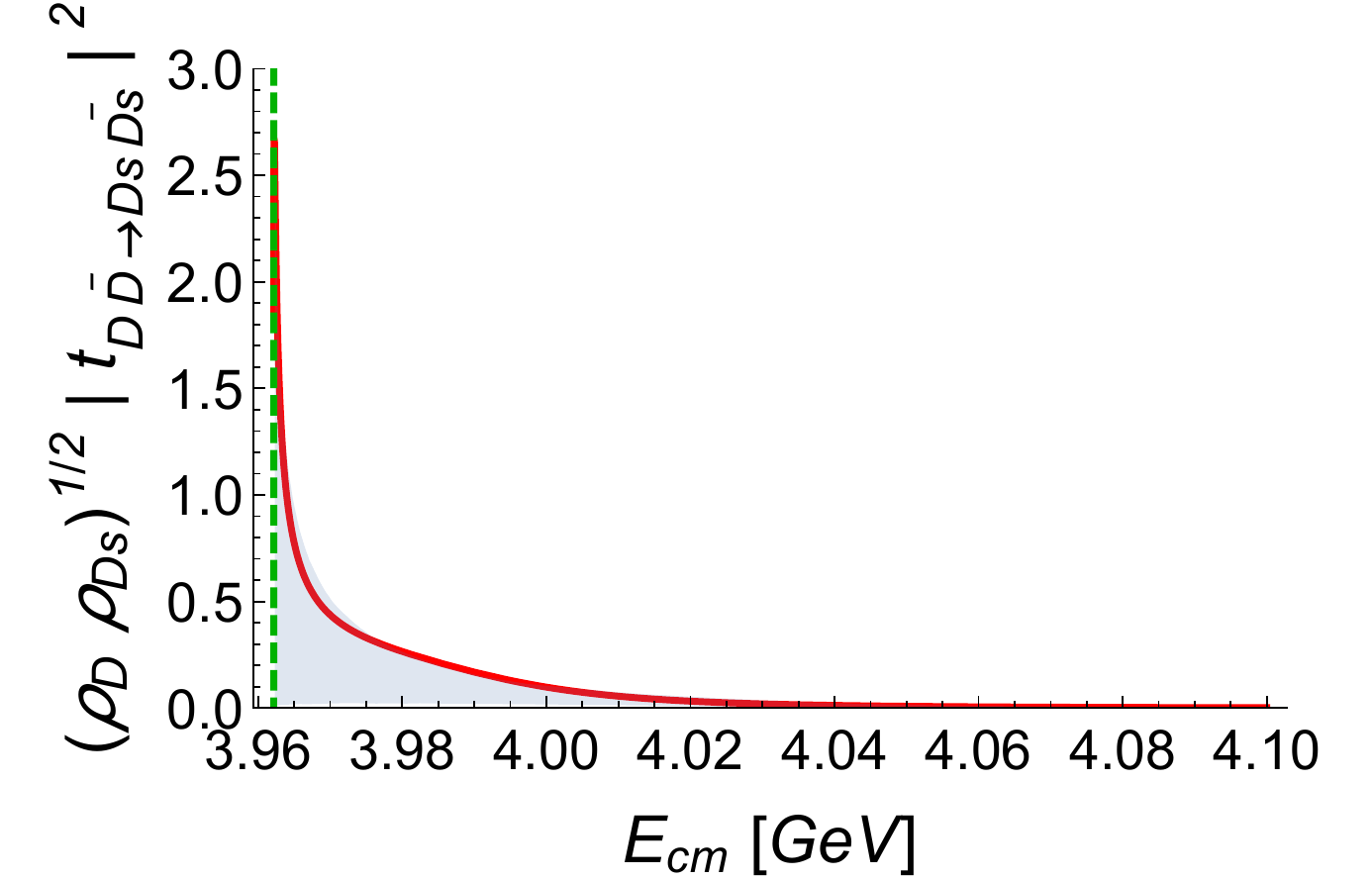}   
		\caption{   Coupled channel $D\bar D$, $D_s\bar D_s$ scattering  in  partial wave $l\!=\!0$. The quantity    $\sqrt{\rho_i\rho_j} |t_{ij}|^2\propto p|\sigma| $ related to the number of events in experiment  is shown ($\rho=2p/E_{cm}$ and $t$ is the scattering matrix).}
	\label{fig:DD-DsDs-with2-t}
	\end{center}
\end{figure}

\item{\bf{Narrow $\bm{0^{++}}$ state  dominated by $\mathrm{D_s\bar D_s}$ slightly below threshold   and  relation to \small{$\bm{X(3915)}/\bm{\chi_{c0}(3930)}$}}}
          
       The    coupled  $D\bar D-D_s\bar D_s$ scattering gives also an indication for a   narrow $0^{++}$ 
       resonance slightly below the $D_s\bar D_s$  threshold, which shows up as  the sharp increase  at the threshold in Fig.  \ref{fig:DD-DsDs-with2-t}.   
        It has a large coupling to $D_s\bar D_s$ and a
       very small coupling to $D\bar D$, as shown in Fig. 6 of  \cite{Prelovsek:2020eiw}. The latter is
       responsible for its small decay rate to $D\bar D$ and the small total
       width.  
       We compare the resulting resonance parameters with the experimental state $X(3915)/\chi_{c0}(3930)$ that shares similar features  to the state we find
       \begin{align}
\label{chic0DsDs} 
     \mathrm{lat}: \ \ &m-2m_{D_s}=-0.2^{~+0.16}_{~-4.9}~\mathrm{MeV}\ ,\quad \ g=0.10^{~+ 0.21}_{~-0.03}~\mathrm{GeV}~\\
     \mathrm{exp}\ X(3915)/\chi_{c0}(3930): \ & \ m-2m_{D_s}=-14.3  \pm 1.8 ~\mathrm{MeV}~,\enskip g=0.69 \pm 0.07  ~\mathrm{GeV}~.  \nonumber  
     \end{align} 
    The $\chi_{c0}(3930)$ with $J^{PC}=0^{++}$ was  recently discovered in
$D\bar D$ decay by LHCb \cite{chic03930}, while the $X(3915)$ with $J^{PC}\!=\!0^{++}$ or  $2^{++}$ was found in $J/\psi \omega$ decay by Belle  \cite{Uehara:2009tx} and BaBar \cite{Aubert:2007vj,delAmoSanchez:2010jr,Lees:2012xs}: both states are now listed  as the same state in the PDG \cite{pdg}.  It lies high above the $D\bar D$ threshold, so one would naturally expect a broad width if it was a conventional charmonium, given the large phase space available to $D\bar D$ decay. Its narrow experimental width   $\Gamma=18.8\pm 3.5~$MeV can only be
explained if  its decay  to $D\bar D$ is suppressed by some mechanism. 

 Our lattice results  suggest that the $X(3915)/\chi_{c0}(3930)$ charmonium-like state is  likely not a conventional  charmonium $\bar cc$, but owes its existence to a large
interaction in the $D_s\bar D_s$ channel near threshold.  
 It lies just below this threshold  on the lattice  and in experiment. Our results indicate that it   has a very small coupling to $D\bar D$, which explains why its  width is small and its decay to $D\bar D$ 
is suppressed in experiment.    

A state dominated by $\bar cc\bar ss$ in this energy region was found also in phenomenological studies  of molecular  \cite{Dong:2021juy} and diquark-antidiquark  
 \cite{Lebed:2016yvr,Giron:2020qpb,Chen:2017dpy} states. 
 
 \item{\bf{ $\bm{1^{--}}$  state related to  $\bm{\psi(3770)}$}}
 
 The  $\psi(3770)$ related to conventional $\bar cc$ state  $1^3D_1$  appears in experiment just above the $D \bar D$ threshold.  
  In our lattice simulation with  $m_{u/d}>m_{u/d}^{exp}$  and $m_c\gtrsim m_c^{exp}$,   we find   it as a bound   state   in the $D\bar D$ scattering amplitude for partial wave $l=1$  slightly below threshold. The resulting resonance parameters show nice agreement with experiment\footnote{The coupling $g$   is defined in (\ref{references}), while the value of $\sqrt{6\pi} g$ is more commonly  listed for vector resonances. The coupling for the bound state   was extracted using $p^3 \cot\delta_1/E_{cm} = (m^2-E_{cm}^2)/g^2$.  }:
  \begin{align}
  \label{psi3770} 
     \mathrm{lat}: \ \ &m-M_{av}=707\pm 7~\mathrm{MeV}\ ,\quad \qquad \sqrt{6\pi} g=18.9^{~+ 0.8}_{~-0.7}~   \\
     \psi(3770)\ \  \mathrm{exp}: \ & \ m-M_{av}= 704.25\pm 0.35 ~\mathrm{MeV}~,\enskip \sqrt{6\pi} g=18.7 \pm 0.9  ~.  \nonumber
            \end{align}  
            We note that $\psi(3770)$ is found as a resonance for our lighter charm quark mass in \cite{Piemonte:2019cbi} and the corresponding coupling $g$ has a very similar value. 
  
 \item{\bf{ $\bm{3^{--}}$  resonance related to  $\bm{\psi_3(3842)}$}}
 
 The $D\bar D$ scattering with partial wave $l=3$ shows a resonance pole that corresponds to the charmonium with $J^{PC}=3^{--}$: 
  \begin{align}
  \label{psi3} 
     \mathrm{lat}: \ \ &m-M_{av}=754~{^{+4}_{-7}}~\mathrm{MeV}\\
     \psi_3(3842)\ \  \mathrm{exp}: \ & \ m-M_{av}= 773.9\pm 0.2  ~,\enskip \Gamma=2.8 \pm 0.6  ~.  \nonumber
            \end{align} 
            We are not able to determine  its width since there is no $D\bar D$ lattice energy level within the energy region of this narrow resonance. 
            The candidate for conventional charmonium with $J^{PC}=3^{--}$ was discovered recently by LHCb     \cite{LHCb:2019lnr} and the mass is close to our lattice value.   
\end{itemize}

\section{Conclusions}

 Charmonium-like states with isospin zero   are  studied by simulating one-channel and coupled-channel scattering on the lattice.    We find all conventional charmonium resonances and bound states with $J^{PC}=0^{++},~1^{--},~2^{++},~3^{--}$ up to the $D_s\bar D_s$ threshold. In addition, the results suggest the existence of unconventional states just below $D_s\bar D_s$ and $D\bar D$ thresholds,  where the first is likely related to $X(3915)/\chi_{c0}(3930)$, while the second state has not been discovered yet. The study makes several simplifying assumptions  necessary for a first investigation of this coupled-channel system.

 
 \vspace{0.5cm}
 
 {\bf Acknowledgments}
 
We thank G. Bali, V. Baru, T. Gershon, F.-K. Guo, B. H\"orz, D. Johnson,  C.~B.~Lang, R. Molina, J. Nieves, E. Oset, S. Paul, 
A.~Sch\"afer and J. Simeth for useful discussions.  The simulations were performed on the Regensburg iDataCool and Athene2 clusters, and the SFB/TRR 55 QPACE~2 and QPACE~3 machines. 
The work is supported by  DFG    SFB/TRR-55, the
European Union’s Horizon 2020 Research and Innovation programme under
the Marie Sklodowska-Curie grant 813942 (ITN EuroPLEx), the STRONG-2020 project grant  824093,  EU  grant MSCA-IF-EF-ST-744659 (XQCDBaryons) and Slovenian Research Agency  grants P1-0035 and
 J1-8137.


\end{document}